\newcommand{\lredcell}{\color{red!60!black}}
\newcommand{\lgreencell}{\color{green!50!black}}
\newcommand{\lbluecell}{\color{blue!60!black}}
\documentclass[10pt,twocolumn,letterpaper]{article}
\usepackage{colortbl}
\definecolor{shadegray}{RGB}{230,230,230}
\definecolor{lightgray}{RGB}{230,230,230}
\usepackage{multirow}
\usepackage{amsmath}
 \usepackage[pagenumbers]{cvpr} 

\usepackage[dvipsnames]{xcolor}

\definecolor{cvprblue}{rgb}{0.21,0.49,0.74}
\usepackage[pagebackref,breaklinks,colorlinks,citecolor=cvprblue]{hyperref}

\def\paperID{16533} 
\def\confName{CVPR}
\def\confYear{2025}

\title{Image Quality Assessment: Exploring Regional Heterogeneity via Response of Adaptive Multiple Quality Factors in Dictionary Space}

\author{Xuting Lan, Mingliang Zhou, Jielu Yan, Xuekai Wei, Yueting Huang, Zhaowei Shang, Huayan Pu 
}

\begin{document}
\maketitle

\begin{abstract}
	Given that the factors influencing image quality vary significantly with scene, content, and distortion type, particularly in the context of regional heterogeneity, we propose an adaptive multi-quality factor (AMqF) framework to represent image quality in a dictionary space, enabling the precise capture of quality features in non-uniformly distorted regions. By designing an adapter, the framework can flexibly decompose quality factors (such as brightness, structure, contrast, etc.) that best align with human visual perception and quantify them into discrete visual words. These visual words respond to the constructed dictionary basis vector, and by obtaining the corresponding coordinate vectors, we can measure visual similarity. Our method offers two key contributions. First, an adaptive mechanism that extracts and decomposes quality factors according to human visual perception principles enhances their representation ability through reconstruction constraints. Second, the construction of a comprehensive and discriminative dictionary space and basis vector allows quality factors to respond effectively to the dictionary basis vector and capture non-uniform distortion patterns in images, significantly improving the accuracy of visual similarity measurement. The experimental results demonstrate that the proposed method outperforms existing state-of-the-art approaches in handling various types of distorted images. The source code is available at https://anonymous.4open.science/r/AMqF-44B2.
\end{abstract}

\section{Introduction}
\label{sec:intro}
\begin{figure*}[!t]
	\centering
	\setlength{\abovecaptionskip}{0.cm}
	\includegraphics[width=1.0\textwidth]{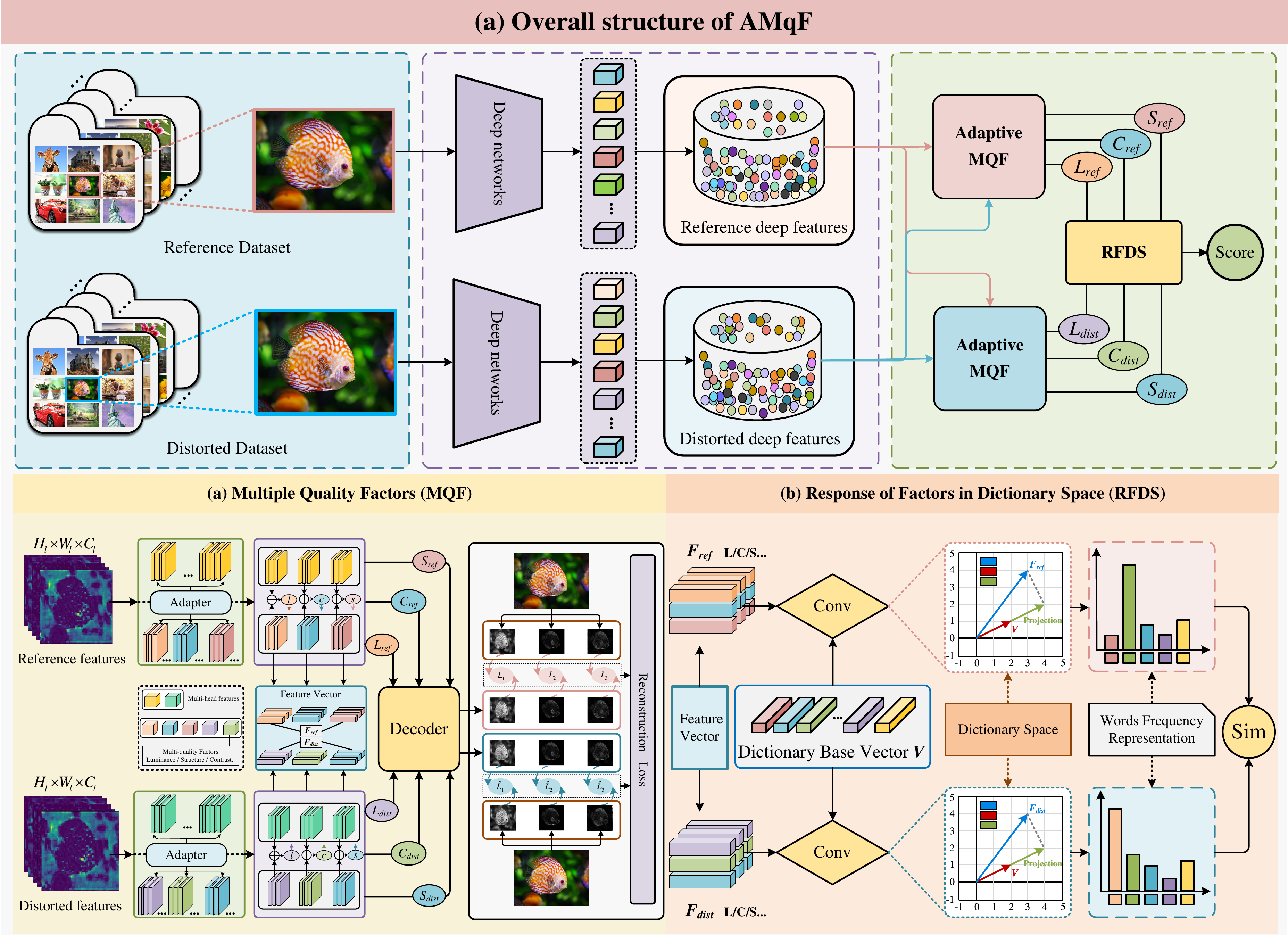}
	
	\caption{Framework of our proposed method (AMqF).}
	\label{method}
\end{figure*}

In the field of digital image processing and multimedia applications, image quality assessment (IQA) plays a crucial role as a core task in evaluating the visual quality of images. Image quality not only directly affects the user's viewing experience but also profoundly affects the subsequent image processing and analysis stages. Full-reference image quality assessment (FR-IQA) \cite{shen, liao2024image, transformer4,AHIQ,CLIB-FIQA,cao2022incorporating,EEGPMAI} is a widely used approach that evaluates the visual quality of distorted images by comparing the differences or similarities between reference and distorted images. The FR-IQA relies on an accurate reference image as a quality benchmark, offering an objective evaluation for various types of distortions. Traditional FR-IQA metrics such as the mean squared error (MSE), mean absolute error (MAE), and peak signal-to-noise ratio (PSNR) exhibit good computational efficiency when measuring pixel differences between images, but they have limitations in providing results that are consistent with human visual perception. As image processing demands continue to grow, more complex distortion types and application scenarios pose higher requirements for IQA, prompting researchers to explore more advanced metrics to simulate the human visual system (HVS) more accurately.

In recent years, the rapid development of deep learning has led to significant advancements in the IQA field, particularly in the FR-IQA task, where deep learning methods have shown great potential \cite{Swiniqa, bakurov2023full,Musiq, ding2021comparison, hammou2021egb,Topiq}. Researchers have proposed various FR-IQA paradigms based on pretrained convolutional neural networks (CNNs) \cite{VGG,resnet50} by extracting multi-level semantic features from both reference and distorted images to measure their similarity or difference \cite{kim2020dynamic, tang2023unifying, cheon2021perceptual}. Ding \textit{et al.} \cite{DISTS} proposed a method called DISTS that combines structure and texture similarity to assess the difference between reference and distorted images. Building on this, Ding \textit{et al.} \cite{ADISTS} introduced a locally adaptive structure and texture similarity framework (A-DISTS), which models the local structure and texture features of images adaptively to improve similarity measurement accuracy. Zhang \textit{et al.} \cite{LPIPS} proposed a perceptual metric based on deep features by computing the distance between reference and distorted images in deep feature space to reflect their perceptual differences. Moreover, Liao \textit{et al.} \cite{liao2022deepwsd} further utilized the Wasserstein distance to measure the difference between reference and distorted images in deep feature space, enhancing the performance of IQA.

Although many deep feature-based FR-IQA methods have been proposed, these methods still fall short when faced with long-standing challenges. On the one hand, handling regional heterogeneity and non-uniform distortion remains difficult. Many classical IQA methods, such as SSIM \cite{wang2004image} and PSNR, assume that distortions in an image are globally uniform. However, in practical scenarios, images often exhibit non-uniform distortions, where certain regions may be more severely distorted than others. On the other hand, traditional methods overlook the nonlinear response of the HVS to different features. The impact of different quality factors on visual perception is not constant but varies significantly depending on the scene, content, and type of distortion. To address these issues, we propose an adaptive multi-quality factor (AMqF) response strategy in a dictionary space, which adaptively selects quality factors that align closely with human visual perception. By quantizing the quality factors into discrete visual words and responding with the dictionary basis vectors within a comprehensive dictionary space, we capture the similarity between quality factors that reflect regional heterogeneity in images. Our framework can more accurately measure image quality and is suitable for assessing images with different types of distortion. Our main contributions are as follows:
\begin{itemize}
	\item Given the systematic differences in how the human eye perceives heterogeneous visual features across regions, we propose an adaptive framework (AMqF) designed to obtain quality factors that closely align with human perception. Specifically, we design an adapter mechanism that can adaptively decompose quality factors (such as brightness, structure, contrast, etc.) from deep features in accordance with human visual quality perception. To further enhance the expressiveness of these quality factors, we introduce a single-channel image reconstruction decoder aimed at strengthening the representation of key quality factors in the IQA, thereby achieving more accurate quality scoring.
	\item To effectively capture distortion patterns in non-uniform regions, we quantize deep features from the image into discrete visual words and construct a complete and distinctive dictionary space encompassing visual words representing different visual attributes. These visual words form a dictionary basis vector, allowing each quality factor to respond within this basis vector space to obtain coordinate vectors. By calculating the similarity between these coordinate vectors, we can improve the precision of the visual similarity measurement.
	\item To cope with different types of distorted images, the framework (AMqF) we design can flexibly adapt to variations in different influencing factors, demonstrating strong robustness and generality and ensuring efficiency and accuracy in various application scenarios.
\end{itemize}

\section{Related Work}
\label{sec:formatting}

\subsection{Image Quality Assessment}
In the field of IQA, numerous classical FR metrics have been developed to simulate the HVS perception of image distortions. These methods emphasize different aspects of image quality, forming a complementary evaluation framework. One of the most representative methods is the structural similarity (SSIM) index, proposed by Wang \textit{et al.} \cite{wang2004image}. This method calculates image similarity by comparing brightness, contrast, and structural information and has become the foundation for many subsequent approaches. Building on this, Wang \textit{et al.} further introduced multiscale SSIM (MS-SSIM) \cite{MS-SSIM}, which evaluates image quality across multiple scales to better capture detail variations at different resolutions. Additionally, Wu \textit{et al.} \cite{VIF} proposed the visual information fidelity (VIF) index from an information-theoretic perspective. This method, which is based on natural scene statistics and the HVS, measures the amount of visual information preserved in distorted images, emphasizing information fidelity by comparing them with the reference image. To simulate human perception of different types of image distortions, Larson \textit{et al.} \cite{CSIQMAD} introduced the most apparent distortion (MAD) model, which combines models of both subtle and obvious distortions, providing consistent evaluations across various distortion levels. Zhang \textit{et al.} \cite{FSIM} subsequently proposed the feature similarity index (FSIM), which focuses on perceptual features via phase congruency and gradient magnitude. Zhang \textit{et al.} \cite{VSI} also introduced the visual saliency-induced index (VSI), which evaluates image quality by considering visual saliency, emphasizing regions that attract attention. The collective development of these methods has provided multidimensional analytical tools for image quality assessment, advancing image processing and compression algorithms.

\subsection{Deep feature-based Image Quality Assessment}
In recent years, deep learning methods have been widely applied to FR-IQA, especially in achieving breakthroughs in perceptual evaluation on the basis of the distributional similarity of feature representations. Wang \textit{et al.} \cite{wang2005reduced} were the first to use the Kullback‒Leibler divergence (KLD) as a distribution metric to compare the similarity between histograms, applying it to reduced-reference image quality assessment (RR-IQA). By measuring the difference between two probability distributions, KLD successfully reveals the degree of image distortion and effectively simulates the HVS's sensitivity to changes in image information. Furthermore, Liu \textit{et al.} \cite{liu2016perceptual} extended this approach by introducing Jensen–Shannon divergence (JSD) for assessing contrast distortion. Compared with KLD, JSD is more robust when handling larger distribution differences, effectively addressing the asymmetry problem in contrast distortion evaluation. With the advancement of deep learning technologies, Zhang \textit{et al.} \cite{LPIPS} proposed an evaluation method (LPIPS) based on deep neural networks (DNNs), which extracts deep features from images and predicts image quality scores by comparing feature distributions. Lao \textit{et al.} \cite{AHIQ} introduced AHIQ, a hybrid network that combines vision transformers (ViT) and CNNs to improve IQA, particularly for GAN-generated distortions. The work of Delbracio \textit{et al.} \cite{delbracio2021projected} further expanded the application of distribution metrics by projecting image features into a high-dimensional space and using the Wasserstein distance (WSD) to compare the distribution differences between enhanced images and target images. Liao \textit{et al.} \cite{liao2022deepwsd} applied WSD to the FR-IQA framework and proposed a quality evaluation method. By matching distributions in the deep feature space, this method achieves more robust quality predictions. Subsequently, Liao \textit{et al.} \cite{liao2024image} developed a training-free FR-IQA framework that introduced multiple perceptual distance metrics, including JSD, WSD, and KLD, to compare the distributions of deep features. This demonstrated the wide applicability and efficiency of distribution-based metrics in IQA tasks. This framework provides a training-free solution for IQA tasks, flexibly adapting to different distortion types and task scenarios, further advancing the development of IQA methods on the basis of distribution metrics.

Although these methods have shown promising results, images in real-world scenarios often exhibit non-uniform distortions, with certain regions being more severely affected than others. Consequently, these approaches struggle to capture this asymmetry when measuring differences in deep features between reference and distorted images, overlooking both regional heterogeneity and the HVS's nonlinear sensitivity to various image features. To address these limitations in existing IQA methods when handling complex perceptual tasks, we propose an adaptive multi-quality factor (AMqF) framework expressed within a dictionary space. This framework dynamically selects quality factors that align with human perceptual sensitivity and maps them into the dictionary space to capture subtle quality variations across non-uniformly distorted regions. By effectively responding to regional heterogeneity and distortion patterns, our method achieves a refined and perceptual measurement of visual similarity.
\section{Problem Formulation}
Assuming that we have a set of reference images \(I_r\) and distorted images \(I_d\), our goal is to explore an effective evaluation method to quantify the perceptual quality of the distorted images relative to the reference images, ensuring that it aligns with human visual perception. The perceptual quality score can be expressed as follows:
\begin{equation}
	\begin{aligned}
		Q = D_{s}(\phi(I_r), \phi(I_d))
	\end{aligned}
\end{equation}
where \(\phi(I_r)\) and \(\phi(I_d)\) represent the deep features by the deep network \(\phi\) of the reference and distorted images, respectively. \(D_{s}\) denotes the similarity measurement, which measures the difference between the deep features.

In the task of FR-IQA, we focus on the non-uniformity and heterogeneity of distortions across different regions of an image. To adapt to these variations, we emphasize quality factors such as luminance, contrast, and structure, which are the most sensitive to HVS. By projecting multiple quality factors onto the response of dictionary basis vectors in a comprehensive dictionary space, we are able to capture the response patterns of these quality factors that align with perceptual regularities. This approach allows us to measure the visual similarity between images in a manner that is more consistent with perceptual quality. The process is described as follows:
\begin{equation}
	\begin{aligned}
		\min _{\theta}\left[\mathcal{L}\left(\sum_{i=1}^{N} D_{\mathcal{D}\left\lbrace l, c, s\right\rbrace }\left(\phi_{\theta}\left(I_{r}^{i}\right), \phi_{\theta}\left(I_{d}^{i}\right)\right)\right), \mathrm{MOS}\right]
	\end{aligned}
\end{equation}
where \(\theta\) represents the parameters of the deep network \(\phi\), \(\{l, c, s\}\) denotes the quality factors (luminance, contrast, structure, etc.), and \(D_{\mathcal{D}\{l, c, s\}}(\phi_{\theta}(I_r^i), \phi_{\theta}(I_d^i))\) indicates the quality score of the visual features in the dictionary space \(\mathcal{D}\), constrained by the loss function \(\mathcal{L}\).

\section{Methodology}
\subsection{Framework}
To explore an FR-IQA framework capable of handling regional heterogeneity and non-uniform distortions, we propose an adaptive multi-quality factor (AMqF) framework that represents image quality in dictionary space to precisely capture the quality characteristics of non-uniformly distorted regions, as shown in Figure \ref{method}. First, we extract deep features from the reference and distorted images via a deep network. Then, an adapter we designed adaptively decomposes these deep features into multiple quality factors (luminance, contrast, structure, etc.). Given that the human eye is more sensitive to high-saliency features, we enhance each quality factor, extracting features closely aligned with visual perception as visual features. To further strengthen the representation of these visual features, we also design a decoder to reconstruct single-channel quality factor images, ensuring that the visual features effectively reflect the characteristics of the HVS. Next, we construct basis vectors in the dictionary space, quantizing the deep features into visual words to capture the response of each quality factor within the visual dictionary basis vectors. This approach enables the model to calculate the coordinate vectors of features from both distorted and reference images on the dictionary basis vectors. Additionally, this forms a distribution of quality factor responses, serving as the global feature representation of the image in the context of the dictionary basis. Finally, by measuring the visual similarity between these coordinate vectors, we assess image quality. Our proposed AMqF framework effectively reflects distortion variations across different regions, demonstrating significant advantages.

\subsection{Adaptive Multi-Quality Factors}
In the field of IQA, the perceived quality of an image is influenced by multiple factors, with luminance, contrast, and structure playing key roles. These factors align closely with the HVS in terms of quality perception. Furthermore, different types of distortions may affect these factors in various ways. On the basis of this understanding, we propose an adaptive multi-quality factor encoding–decoding joint learning method to better capture and adapt to the diversity and complexity of image quality, as shown in Figure \ref{method} (a).
Specifically, given a batch of distorted images \(I_{ref} \in \mathbb{R}^{N \times H \times W \times 3} \) and reference images \(I_{dist} \in \mathbb{R}^{N \times H \times W \times 3} \) with a sample size of \(N\), where \(H\) and \(W\) are the height and width of the input images, respectively, and 3 denotes the RGB channels, they are input into a backbone network with a pretrained ResNet50 model as the encoder to obtain the deep features of the last layer \(l \), represented as \(\phi_{l}(I_{ref}) \in \mathbb{R}^{H_{l} \times W_l \times C_l} \) and \(\phi_{l}(I_{dist}) \in \mathbb{R}^{H_{l} \times W_l \times C_l} \).
The quality factors that align closely with human perception, namely, luminance (\(l\)), contrast (\(c\)), structure (\(s\)) and among others, were subsequently adaptively decomposed. Considering the potential for information loss during the decomposition of features, an additional feature expansion operation was performed to better fit the characteristics representing each quality factor. Specifically, the deep features \(\phi_{l}(I_{ref})\) and \(\phi_{l}(I_{dist})\) obtained in the initial stage were adaptively divided into multi-head features \(F_{h} \{I_{ref}, I_{dist}\} \) and the remaining highly sensitive features \(\left\lbrace  L_{ref}, C_{ref}, S_{ref}\right\rbrace  \) and \(\left\lbrace  L_{dist}, C_{dist}, S_{dist} \right\rbrace \), allowing the multi-head features to automatically supplement any missing parts of the visual features. The multi-head features are then added to the remaining features to produce the updated visual features, which are used to represent the image in the dictionary space.

In addition to adaptively capturing multi-quality factor features, we also introduce a mechanism for image reconstruction to make these features closer to human visual perception. Specifically, the quality factors features of the reference and distorted images (luminance, contrast, structure) are each input into a decoder for single-channel image reconstruction. During training, the features of the quality factors (luminance, contrast, and structure) of both the original distorted image and the reference image are extracted to constrain the reconstructed image. To achieve this, we design a reconstruction loss function \(\mathcal{L}_{\text{re}} \), combining image gradient loss \(\mathcal{L}_{\text{grad}} \) and intensity loss \(\mathcal{L}_{\text{intensity}} \) to compare the two images before and after reconstruction \( I_1 \) and \( I_2 \). The specific formulation is as follows:
\begin{equation}
	\begin{aligned}
		\mathcal{L}_{\text{re}} &= \mathcal{L}_{\text{grad}} + \mathcal{L}_{\text{intensity}}\\&= \sum_{c \in \{x, y\}} \left\| \nabla_c I_1 - \nabla_c I_2 \right\|_1 + \left\| I_1 - I_2 \right\|_1
	\end{aligned}
\end{equation}
where \(c \) is the subscript representing the gradient direction, corresponding to the vertical \(x \) and horizontal \(y \) directions, \(\nabla_c \) represents the gradient, \(\mathcal{L}_{\text{grad}} \) measures the structural similarity, and \(\mathcal{L}_{\text{intensity}} \) measures the pixel-level luminance difference.

\begin{table*}[!t]
	\centering
	\setlength{\tabcolsep}{12.2pt}
	\renewcommand\arraystretch{0.7}
	
	\begin{tabular}{lcccccccccc}
		\toprule [1.5pt]
		\multirow{2}{*}{Method} & \multicolumn{2}{c}{LIVE \cite{LIVE}} & \multicolumn{2}{c}{CSIQ \cite{CSIQMAD}} & \multicolumn{2}{c}{TID2013 \cite{TID2013}} & \multicolumn{2}{c}{KADID-10k \cite{kadid}} \\
		\cmidrule(lr){2-3} \cmidrule(lr){4-5} \cmidrule(lr){6-7} \cmidrule(lr){8-9}
		& PLCC  & SRCC  & PLCC  & SRCC  & PLCC  & SRCC  & PLCC  & SRCC  \\
		\midrule
		PSNR                    & 0.791 & 0.807 & 0.781 & 0.801 & 0.663 & 0.686 & 0.667 & 0.672 \\
		\rowcolor{shadegray}	SSIM \cite{wang2004image}               & 0.847 & 0.851 & 0.819 & 0.832 & 0.665 & 0.627 & 0.780 & 0.778 \\
		MS-SSIM \cite{MS-SSIM}           & 0.886 & 0.903 & 0.864 & 0.879 & 0.785 & 0.729 & 0.835 & 0.834 \\
		\rowcolor{shadegray}	VIF \cite{VIF}             & 0.948 & 0.952 & 0.898 & 0.899 & 0.771 & 0.677 & 0.676 & 0.669 \\
		MAD     \cite{CSIQMAD}           & 0.904 & 0.907 & 0.934 & 0.932 & 0.803 & 0.773 & 0.829 & 0.827 \\
		\rowcolor{shadegray}	FSIM     \cite{FSIM}           & 0.910 & 0.920 & 0.902 & 0.915 & 0.876 & 0.851 & 0.850 & 0.850 \\
		VSI    \cite{VSI}           & 0.877 & 0.899 & 0.912 & 0.928 & 0.898 & 0.894 & 0.875 & 0.876 \\
		\rowcolor{shadegray}	GMSD     \cite{GMSD}          & 0.909 & 0.910 & 0.938 & 0.939 & 0.858 & 0.804 & 0.847 & 0.846 \\
		NLPD  \cite{NLPD}            & 0.882 & 0.889 & 0.913 & 0.925 & 0.832 & 0.799 & 0.819 & 0.820 \\
		
		\hline
		\rowcolor{shadegray}WaDIQaM-FR  \cite{WaDIQaM-FR}       & 0.980 & 0.970 &  0.967 &  0.962 &   0.946 &   0.940 & \lbluecell \bf 0.935 & \lgreencell \bf 0.931 \\	
		DeepQA  \cite{DeepQA}    & 0.982 & \lgreencell \bf 0.981 &  0.965 & 0.961 &  0.947 & 0.939 &   0.910 & \lbluecell \bf 0.912 \\
		\rowcolor{shadegray}	PieAPP   \cite{pieapp} & \lgreencell \bf 0.986 & 0.977 &\lbluecell \bf 0.975 &   0.973 &  0.946 &  0.945 & 0.832 & 0.830 \\
		DeepFL-IQA \cite{DeepFL-IQA}   &0.978 &0.972&0.946 &0.930&0.876 &0.858&0.938 &0.936\\
		\rowcolor{shadegray}	JND-SalCAR \cite{JND-SalCAR} & \lredcell \bf 0.987 &\lredcell \bf 0.984 & \lgreencell \bf 0.977 & \lgreencell \bf 0.976&\lbluecell \bf 0.956 &\lbluecell \bf 0.949& \lgreencell \bf 0.960 &0.959\\
		LPIPS-VGG    \cite{LPIPS}         &0.978 &0.972&0.970 &0.967&0.944 &0.936&- &-\\
		\rowcolor{shadegray}	DISTS    \cite{DISTS}         &0.954 &0.954&0.928 &0.929&0.855 &0.830&0.886 &0.887\\
		A-DISTS		\cite{ADISTS}		&0.955 &0.955 &0.947 &0.941 &0.858 &0.835 &0.892 &0.892 \\
		\rowcolor{shadegray}	DeepWSD \cite{liao2022deepwsd} &0.961 &0.962&0.950&   0.965&0.870&0.874&0.883&0.883	\\
		TOPIQ-FR \cite{Topiq}  &\lbluecell \bf 0.984 &\lredcell \bf 0.984& \lredcell \bf 0.980& \lredcell \bf 0.978 &\lgreencell \bf 0.958& \lgreencell \bf 0.954&-&-	\\
		\hline
		\rowcolor{shadegray}	AMqF (ours)                  & 0.979 &\lbluecell \bf 0.980 &\lbluecell \bf 0.975 & \lbluecell \bf 0.974 & \lredcell \bf 0.968 & \lredcell \bf 0.968 &\lredcell \bf 0.964  &\lredcell \bf 0.961 \\
		AMqF-VGG (ours)                  &\lgreencell \bf \underline{0.986} &\lbluecell \bf 0.980 &\lgreencell \bf \underline{0.977} & \lgreencell \bf \underline{0.976} & \lredcell \bf 0.965 & \lredcell \bf 0.964 &\lredcell \bf \underline{0.967}  &\lredcell \bf \underline{0.965} \\
		\bottomrule [1.5pt]
	\end{tabular}
	\caption{Performance comparison of the proposed AMqF algorithm against state-of-the-art FR-IQA algorithms on benchmark datasets. The best, second, and third results are bolded and highlighted in {\lredcell \bf red}, {\lgreencell \bf green}, and {\lbluecell \bf blue}, respectively. Additionally, the best results for different backbones are underlined.}
	\label{tab:comparison}
\end{table*}

\begin{table}[t]	
	
	\setlength{\tabcolsep}{0.8pt}
	\renewcommand\arraystretch{0.685}
	\begin{tabular}{lcccccc}
		\toprule
		\multirow{2}{*}{Method} & \multicolumn{2}{c}{LIVE \cite{LIVE}} & \multicolumn{2}{c}{CSIQ \cite{CSIQMAD}} & \multicolumn{2}{c}{TID2013 \cite{TID2013}}  \\
		\cmidrule(lr){2-3} \cmidrule(lr){4-5} \cmidrule(lr){6-7}
		& PLCC  & SRCC  & PLCC  & SRCC  & PLCC  & SRCC  \\
		\hline
		PSNR    & 0.865 & 0.873 & 0.786 & 0.809 & 0.677 & 0.687 \\	
		\rowcolor{shadegray} WaDIQaM-FR \cite{WaDIQaM-FR}   & 0.837 & 0.883   & - & -  & 0.741 & 0.698 \\
		RADN \cite{RADN} & 0.878 & 0.905   & - & -  & 0.796 & 0.747 \\
		\rowcolor{shadegray}	AMqF (ours)         & \bf 0.907 & \bf0.926 & \bf 0.866 & \bf0.862 &\bf 0.830 & \bf 0.815 \\
		\toprule
	\end{tabular}
	\vspace{-0.5em}
	\caption{Cross-database comparison results obtained from training on the entire KADID-10k database and performance evaluation for cross-database assessments.}
	\label{tab:cross}
	\centering
\end{table}
\subsection{Response of Factors in Dictionary Space}
We construct a comprehensive and highly discriminative visual dictionary space \(\mathcal{D} \) to address the varying distortion effects across different regions of an image. This space effectively captures visual features through adaptive multi-quality factors, enabling a complete representation of the image content with non-uniform distortions, as shown in Figure \ref{method} (b). In this space, the image is represented as a dictionary of discrete visual words, where the response intensity of each visual word forms the feature vector of the image. To achieve this, we introduce a dictionary basis defined by a learnable parameter matrix \(\mathcal{V} \in \mathbb{R}^{1024 \times 512} \), with dimensions \(N \times D \), where \(N \) denotes the number of visual words in the dictionary, and \(D \) represents the feature dimension of each visual word (consistent with the image feature dimension). The dictionary basis vectors are expressed as follows:
\begin{equation}{\label{bovw}}
	\begin{split}
		\mathbf{V} \in \mathbb{R}^{N \times D}, \quad \mathbf{V}_i \in \mathbb{R}^{D}, \quad i \in \{1, 2, \dots, N\}
	\end{split}
\end{equation}
Initially, the dictionary basis vectors are initialized via a Kaiming normal distribution and dynamically updated during training to adaptively learn feature patterns for various distortion types. Before projecting the adaptive quality factors onto the visual dictionary basis vectors, we apply \(L_2\) normalization to the visual features to eliminate scale effects, ensuring consistency between the feature vectors. To capture the responsiveness of visual features within the dictionary basis vectors, we perform convolution between the normalized feature vectors and the visual dictionary basis vectors. The output represents the response of each quality factor for the reference and distorted images across different spatial regions, reflecting the mapping coordinates of the image's local features in the dictionary basis vectors. The responses of the quality factors are denoted as \(R_{ref}[i, j]\) and \(R_{dist}[i, j]\), capturing regional heterogeneity and distortion patterns within the image. This responsiveness can be described by the following equation:
\begin{equation}
	\begin{split}
		R_k[i, j] = \sum_{c=1}^{C} \sum_{m=1}^{H} \sum_{n=1}^{W} \mathcal{F}[i+m, j+n, c] \cdot v_k[m, n]
	\end{split}
\end{equation}
where \((i, j) \) denotes the position in the response map, \((m, n) \) represents the spatial position of the convolution kernel, and \(R_k[i, j] \) denotes the projection coordinate of the feature vector in the direction of visual word \(v_k \), reflecting the degree of match with different feature patterns within the image. Finally, we use average pooling to aggregate the response map of each visual word under each quality factor into a scalar value, representing the average response intensity of that visual word across the entire image, denoted as \(\mathbf{P}_{dist}[k]\) and \(\mathbf{P}_{ref}[k]\). This forms a distribution of quality factor responses, serving as the global feature representation of the image in the context of the dictionary basis.
\begin{equation}{\label{ score}}
	\begin{split}
		\mathbf{P}_{dist}[k] = \frac{1}{N} \sum_{i=1}^{N} R_k[i, j] \\
		\mathbf{P}_{ref}[k] = \frac{1}{N} \sum_{i=1}^{N} R_k[i, j] \\
	\end{split}
\end{equation}
By computing the cosine similarity between the coordinate vectors of the reference and distorted images, we derive the quality score \(Q \) for the image, which is formulated as follows:
\begin{equation}
	\begin{split}
		Q = \frac{\mathbf{P}_{\text{ref}} \cdot \mathbf{P}_{\text{dist}}}{\|\mathbf{P}_{\text{ref}}\| \|\mathbf{P}_{\text{dist}}\|}
	\end{split}
\end{equation}

To ensure that different features in the feature space are independent from each other and to reduce redundancy and correlation among features, we introduced a decorrelation loss during the training process, described by the following formula:
\begin{equation}{\label{ score}}
	\begin{split}
		\mathcal{L}_{\text{decov}} &= \| C \|_F - \sqrt{\sum_{k=1}^{D} C_{kk}^2 + 1e-6}\\
		\| C \|_F &= \sqrt{\sum_{i=1}^{N} \sum_{j=1}^{N} C_{ij}^2}
	\end{split}
\end{equation}
where \(\| C \|_F \) is the Frobenius norm of the covariance matrix, which represents the square root of the sum of the squares of all the elements, \(D\) is the feature dimension, and \(C_{kk} \) denotes the diagonal elements of the covariance matrix.

\begin{figure*}[t]
	\centering
	\captionsetup[subfloat]{font=scriptsize}
	\renewcommand{\thesubfigure}{\arabic{subfigure}}
	\subfloat[PSNR]{\includegraphics[width=0.257\linewidth]{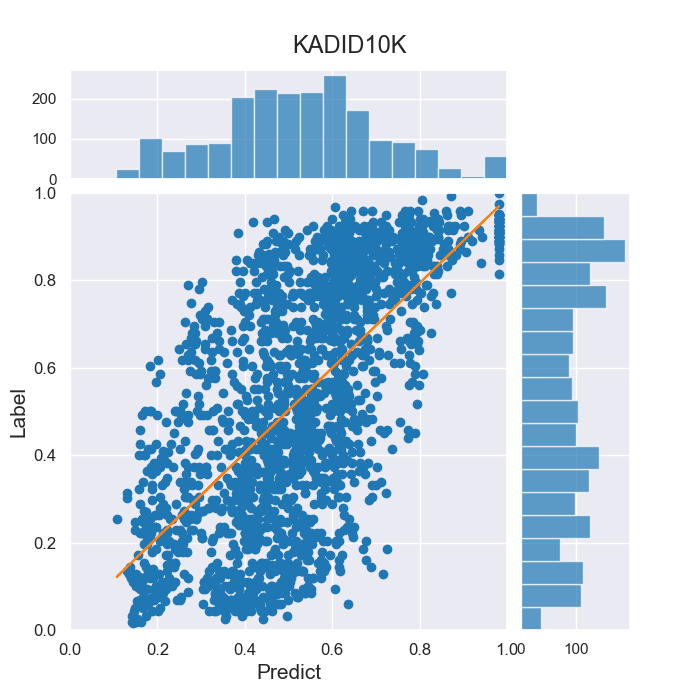}}
	\subfloat[SSIM]{\includegraphics[width=0.257\linewidth]{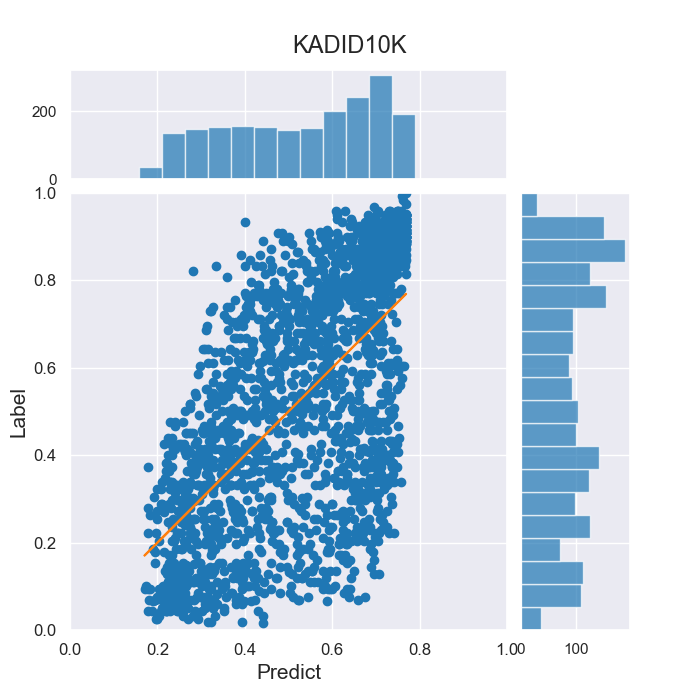}}
	\subfloat[MS-SSIM]{\includegraphics[width=0.257\linewidth]{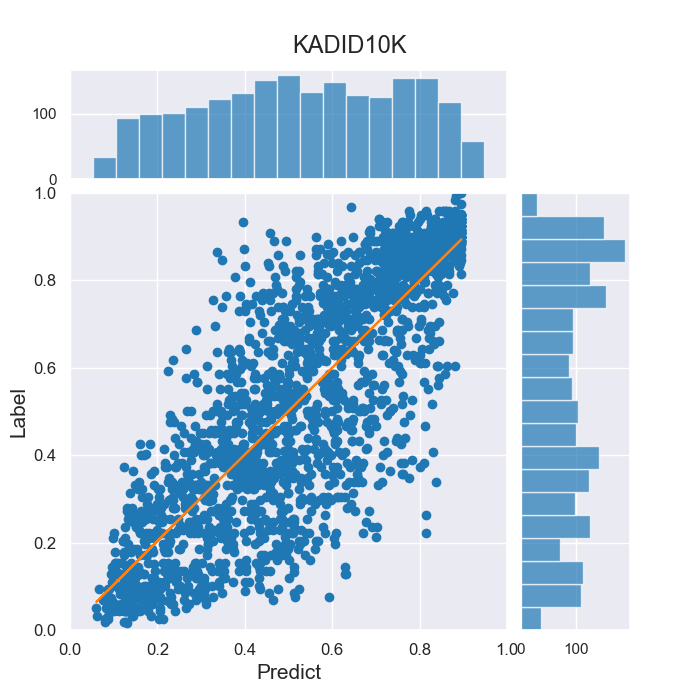}}
	\subfloat[VIF]{\includegraphics[width=0.257\linewidth]{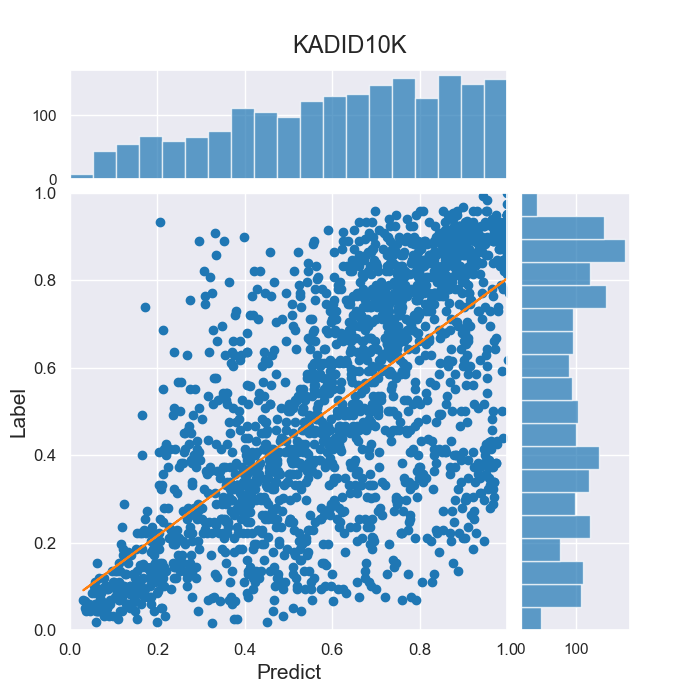}}
	\vspace{-0.3em} \\
	\subfloat[MAD]{\includegraphics[width=0.257\linewidth]{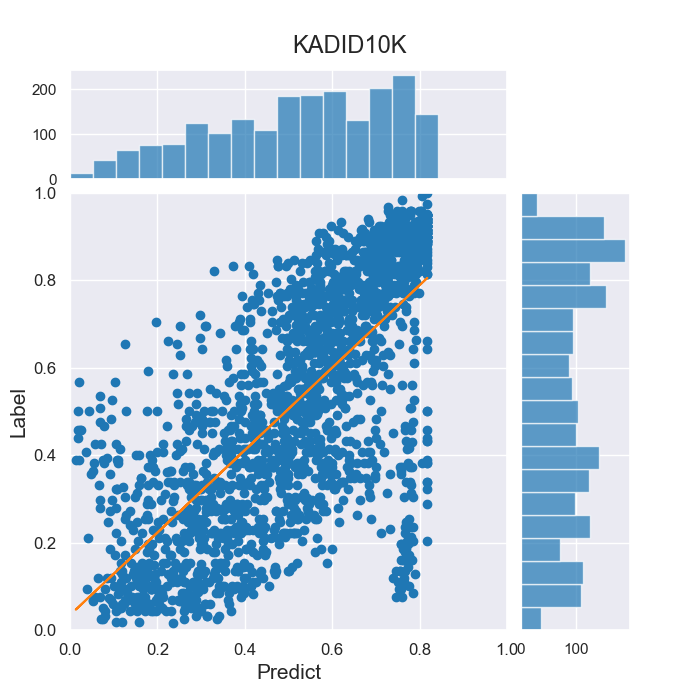}}
	\subfloat[FSIM]{\includegraphics[width=0.257\linewidth]{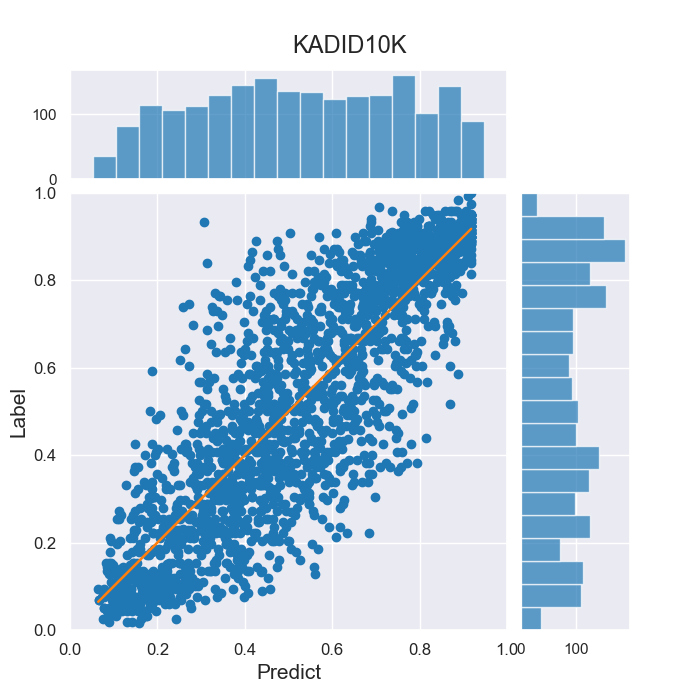}}
	\subfloat[VSI]{\includegraphics[width=0.257\linewidth]{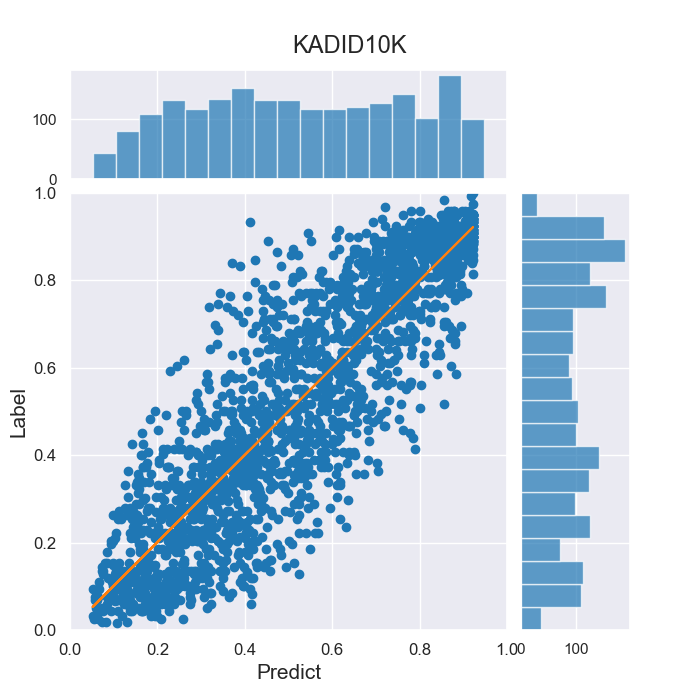}}
	\subfloat[GMSD]{\includegraphics[width=0.257\linewidth]{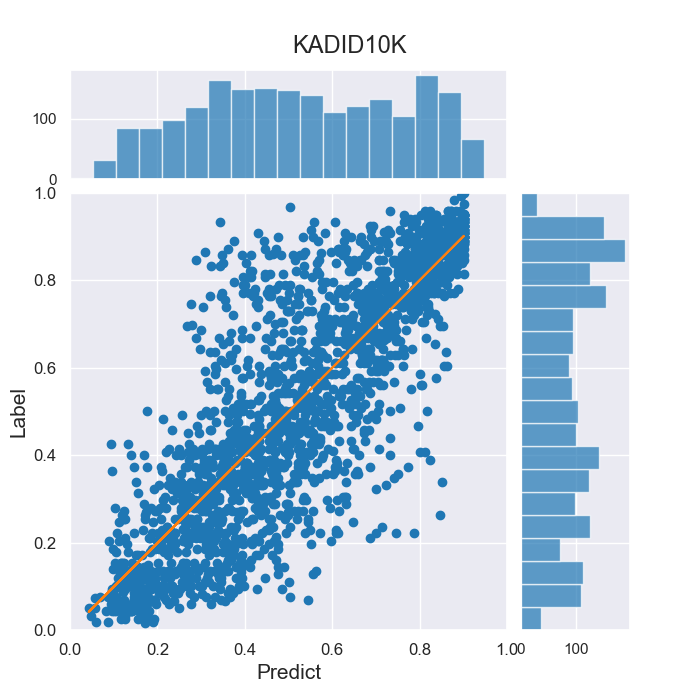}}
	\vspace{-0.3em} \\
	\subfloat[NLPD]{\includegraphics[width=0.257\linewidth]{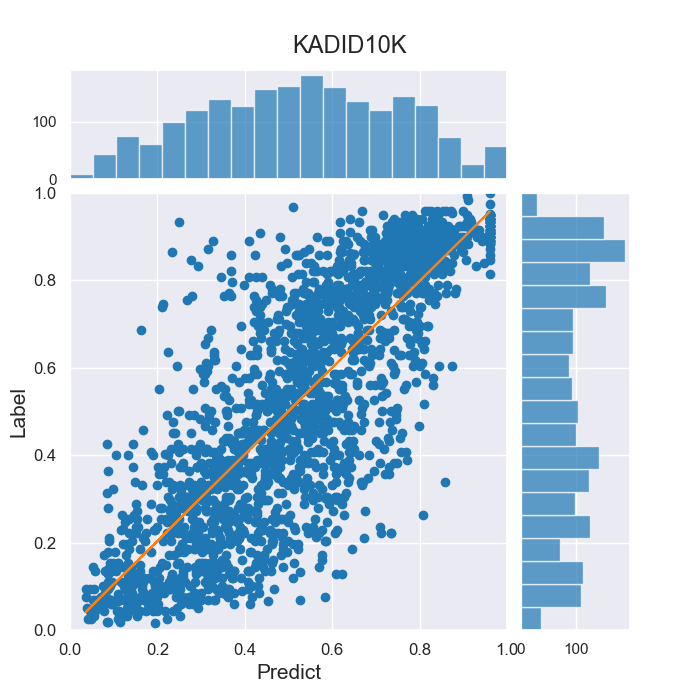}}
	\subfloat[WqDIQaM-FR]{\includegraphics[width=0.257\linewidth]{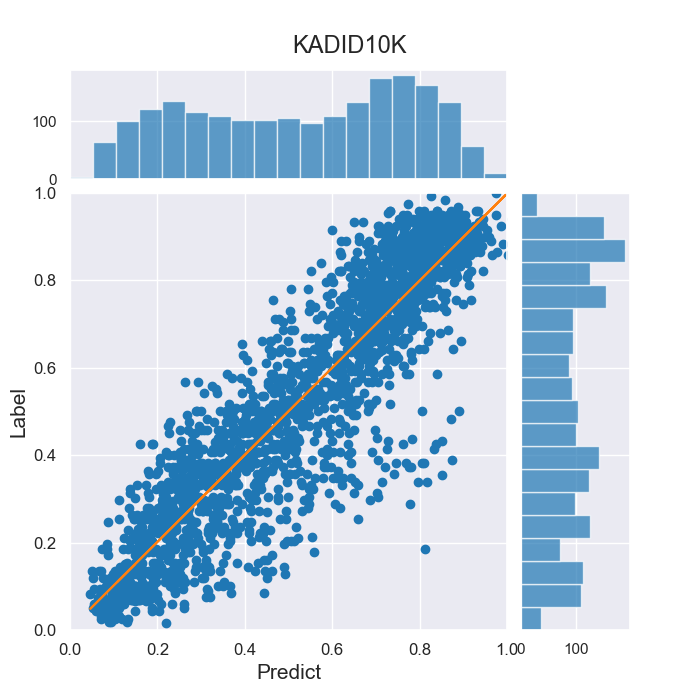}}
	\subfloat[PieAPP]{\includegraphics[width=0.257\linewidth]{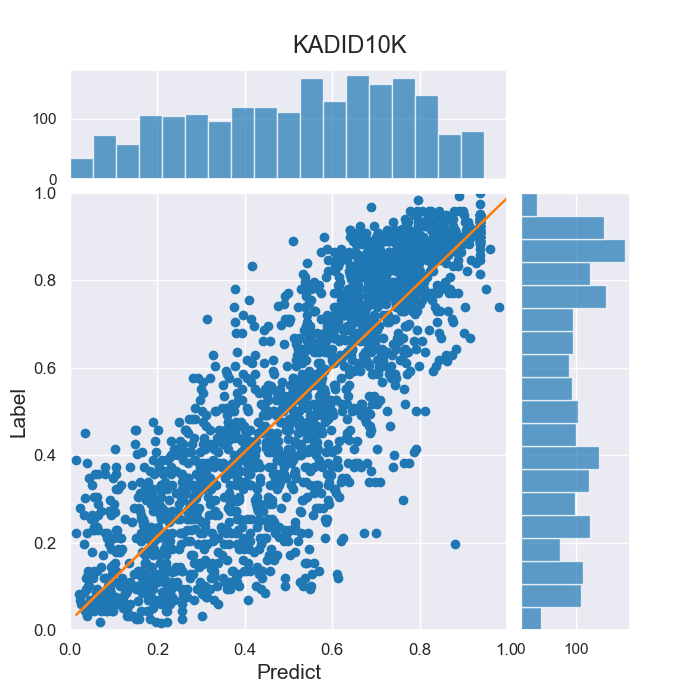}}
	\subfloat[DISTS]{\includegraphics[width=0.257\linewidth]{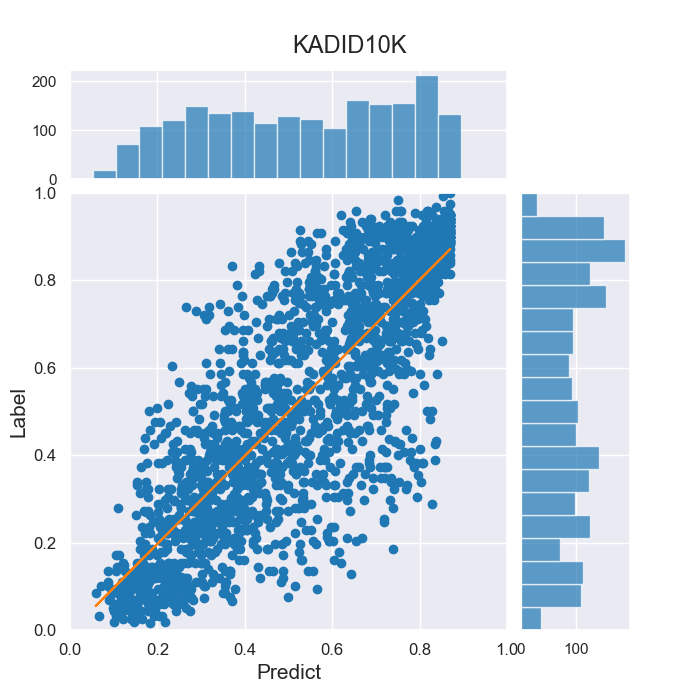}}
	\vspace{-0.2em} \\
	\subfloat[Ours/LIVE]{\includegraphics[width=0.257\linewidth]{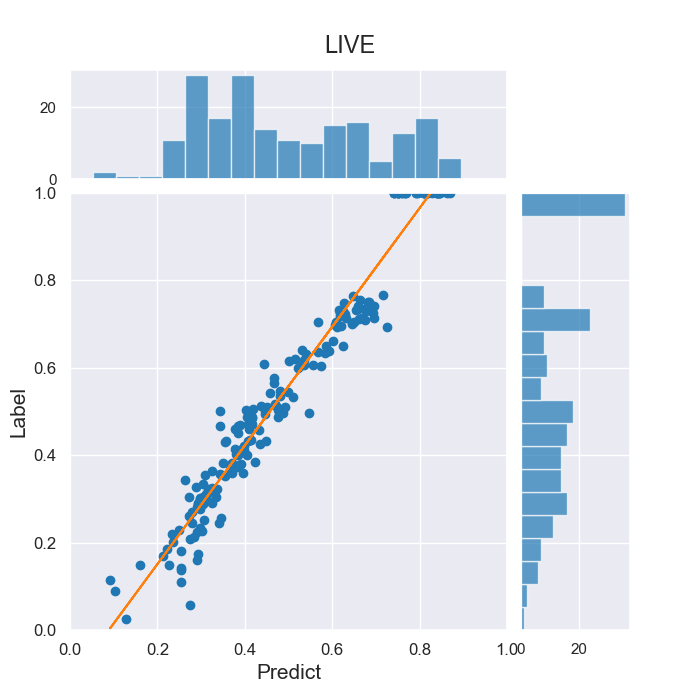}}
	\subfloat[Ours/CSIQ]{\includegraphics[width=0.257\linewidth]{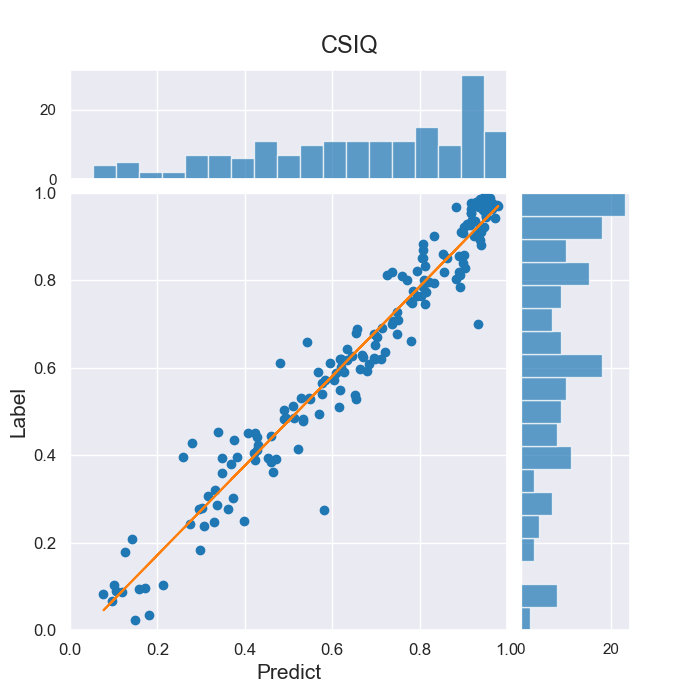}}
	\subfloat[Ours/TID2013]{\includegraphics[width=0.257\linewidth]{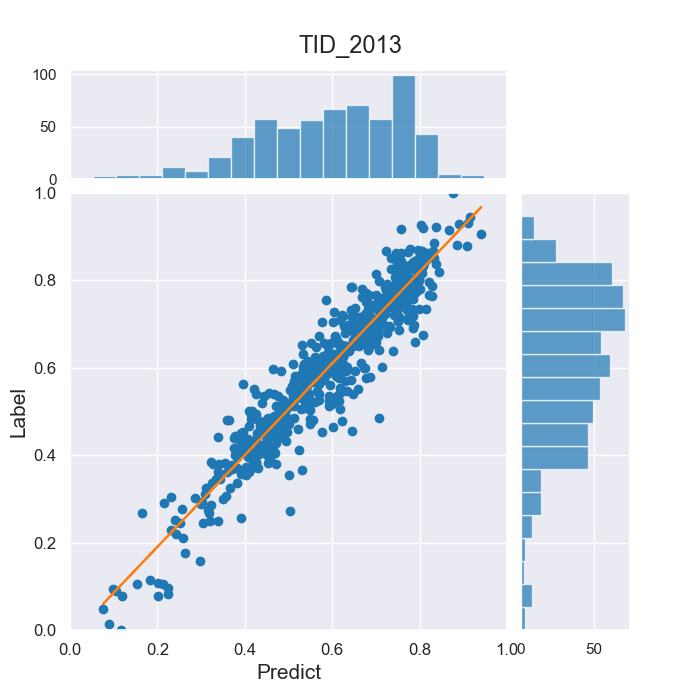}}
	\subfloat[Ours/KADID10K]{\includegraphics[width=0.257\linewidth]{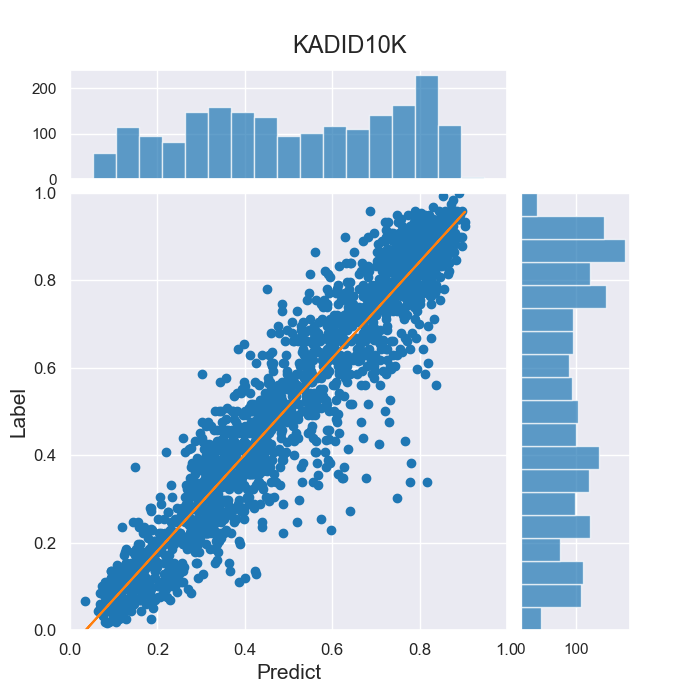}}
	\vspace{-0.1em}
	\caption{Scatter plots of the prediction results for various FR-IQA methods on the KADID-10k \cite{kadid} dataset, as well as scatter plots of our method (AMqF) on the LIVE, CSIQ and TID2013 datasets.}
	\label{visual2}
\end{figure*}

\subsection{Connection to the HVS and existing methods}
Existing FR-IQA methods \cite{LPIPS, delbracio2021projected} focus on comparing deep features. However, these methods overlook the issue of local distortions and the differences in the sensitivity of the human eye to various regions. As a result, they still face limitations when dealing with regional heterogeneity and non-uniform distortions. The proposed method deeply integrates the principles of HVS, particularly with respect to the sensitivity of the human eye to different image features. The HVS responds nonlinearly to changes in various quality factors in an image, such as luminance, contrast, and structure. This sensitivity varies significantly depending on the scene, content, and type of distortion. Unlike traditional methods such as SSIM \cite{wang2004image} and PSNR, which assume global uniformity of distortion, our method breaks this assumption by focusing on local distortions and quality variations. Moreover, recognizing that HVS sensitivity to brightness, contrast, and structure differs, we address this limitation by adaptively selecting quality factors that align with human perception. By decomposing deep features into quality factors that match HVS perceptions and quantifying them into discrete visual words, we can accurately capture the quality characteristics of non-uniformly distorted regions. These visual words respond within the constructed dictionary space, and by obtaining their corresponding coordinate vectors, we measure visual similarity, thereby providing a better reflection of the image's quality.
\begin{table*}[t]
	
	\centering
	
	\setlength{\tabcolsep}{11.5pt}
	\renewcommand\arraystretch{0.5}
	
	\begin{tabular}{lcccccccccc}
		\toprule
		\multirow{2}{*}{Method} & \multicolumn{2}{c}{LIVE \cite{LIVE}} & \multicolumn{2}{c}{CSIQ \cite{CSIQMAD}} & \multicolumn{2}{c}{TID2013 \cite{TID2013}} & \multicolumn{2}{c}{KADID-10k \cite{kadid}} \\
		\cmidrule(lr){2-3} \cmidrule(lr){4-5} \cmidrule(lr){6-7} \cmidrule(lr){8-9}
		& PLCC  & SRCC  & PLCC  & SRCC  & PLCC  & SRCC  & PLCC  & SRCC  \\
		\hline
		AMQF    & 0.978 & 0.979 & 0.971 &\bf 0.979 & \bf 0.968 & 0.966 & 0.947 & 0.958 \\	
		\rowcolor{shadegray} RFDS   & 0.936 & 0.955   & 0.970 & 0.974  & 0.871 & 0.880 & 0.947 & 0.957 \\
		Full model (AMqF)          & \bf 0.979 & \bf0.980 & \bf 0.975 & 0.974 &\bf 0.968 & \bf 0.968 & \bf 0.964 &\bf 0.961 \\
		\toprule
	\end{tabular}
	\vspace{-0.7em}
	\caption{Results of ablation results on four benchmark datasets.}
	\label{tab:ablation}
\end{table*}
\section{Experiments}
\subsection{Experimental setups}
We evaluated the performance of our proposed method via four publicly available datasets, including the LIVE \cite{LIVE},
CSIQ \cite{CSIQMAD}, and TID2013 \cite{TID2013}.
Additionally, to ensure a diverse set of distortion types, we included the large-scale artificially distorted IQA dataset KADID-10k \cite{kadid}.
For evaluation metrics, we selected the Spearman rank order correlation coefficient (SROCC) \cite{SROCC} and the Pearson linear correlation coefficient (PLCC) \cite{PLCC} to compare the performance of our proposed adaptive multi-quality factor (AMqF) method with that of other FR-IQA methods. All the experiments were conducted on a single NVIDIA GeForce RTX 4090 using PyTorch.
Specifically, during the image preprocessing stage, all the input images were randomly cropped to a size of 224$\times$224$\times$3 pixels for input into the AMqF framework.
\subsection{Experimental Results}
\subsubsection{Comparison with the State-of-the-Art Method}
In our study, to comprehensively evaluate the effectiveness of the proposed AMqF framework, we conducted performance comparisons against various classic 19 IQA methods on standard datasets.
These methods include PSNR, SSIM \cite{wang2004image}, MS-SSIM \cite{MS-SSIM}, VSI \cite{VSI}, VIF \cite{VIF}, FSIM \cite{FSIM}, GMSD \cite{GMSD}, NLPD \cite{NLPD}, WaDIQaM-FR \cite{WaDIQaM-FR}, PieAPP \cite{pieapp}, MAD \cite{CSIQMAD}, DISTS \cite{DISTS}, ADISTS \cite{ADISTS}, DeepWSD \cite{liao2022deepwsd}, DeepQA \cite{DeepQA}, DeepFL-IQA \cite{DeepFL-IQA}, JND-SalCAR \cite{JND-SalCAR}, LPIPS-VGG \cite{LPIPS}, and TOPIQ \cite{Topiq}.

Table \ref{tab:comparison} shows a performance comparison between the proposed AMqF framework and current state-of-the-art FR-IQA algorithms across four IQA datasets. Specifically, we selected nine traditional methods and ten deep learning-based methods for benchmarking. The results show that our approach achieves superior overall performance. In particular, it achieves the best results on the TID2013 and KADID-10k. Although it did not achieve the highest performance on the LIVE and CSIQ datasets, it still maintained performance within the top three. This also indicates that our framework exhibits strong adaptability across various distortion scenarios, demonstrating its robustness and effectiveness in handling complex, real-world non-uniform distortions.

To further evaluate the robustness of our method across different datasets, we conducted cross-database experiments to demonstrate the performance of our approach compared with competing methods. Specifically, we trained on the complete KADID-10k database and tested it on several other datasets without any fine-tuning or parameter adaptation. Table \ref{tab:cross} illustrates the comparison results between our proposed method and typical methods. The results indicate that the model trained on KADID-10k performs exceptionally well in cross-database evaluations, particularly in handling complex distortions, highlighting its superior generalization ability and further validating the strong versatility of our method.

To further evaluate the performance of our method, we visualized the true scores and corresponding predicted scores for 12 classic FR-IQA methods. Figure \ref{visual2} displays scatter plots of the prediction results for these methods alongside our proposed AMqF method on the KADID-10k. The results indicate that the AMqF method not only shows a high correlation between the predicted scores and true scores on the KADID-10k dataset but also demonstrates this correlation across three other datasets. This further validates that AMqF achieves outstanding predictive performance across datasets with various distortion types.
\subsection{Ablation Study and Analysis}
To verify the effectiveness of each component in our proposed method and examine its impact on overall performance, we conducted ablation experiments. Our model consists of two parts: adaptive multiple quality factors (AMQF) and the response of factors in dictionary space (RFDS). In the experiments, we removed each component individually to assess its independent contribution. Table \ref{tab:ablation} lists the performance variations after removing AMQF and RFDS. The results show that removing AMQF leads to a decline in model performance, indicating that the RFDS branch plays a crucial role in capturing regional heterogeneity within images. Removing RFDS causes a significant performance drop across the four datasets, further demonstrating that AMQF enhances overall performance by learning the most sensitive quality factors for human vision and embedding their correlations into the model. When both AMQF and RFDS components are combined, our method achieves optimal performance, highlighting the indispensability and complementarity of the two parts. They not only improve the model’s ability to assess image quality individually but also enhance its robustness and adaptability through their synergistic interaction.

\section{Conclusion}
In this paper, we proposed the AMqF framework, which represents image quality in a dictionary space and deeply analyses the nonlinear sensitivity of the HVS to different image features. First, on the basis of the sensitivity of HVS, we adaptively decompose quality factors and enhance their representation through single-channel image reconstruction, which better accommodates local distortions while ensuring close alignment with human visual perception. Second, by constructing a vector basis in the dictionary space, we project the adaptive quality factors into the dictionary space and extract their coordinate vectors in the basis vectors, capturing regional heterogeneity and distortion patterns, thus enabling precise measurement of visual similarity in regions with non-uniform distortions.

{
	\small
	\bibliographystyle{ieeenat_fullname}
	\bibliography{main}
}

\end{document}